\documentclass[12pt,preprint]{aastex}
\shorttitle{Ellipticals from Self-Consistent Simulations}
\shortauthors{S\'aiz et al.}

\received{2003 September 5}
\begin{document}
 
\title{Ellipticals at $z=0$ from Self-Consistent Hydrodynamical
Simulations: Comparison with SDSS Structural and Kinematical Data
}
 
\author{A. S\'aiz$^1$,   R. Dom\'{\i}nguez-Tenreiro
\footnote{ Dpt. F\'{\i}sica Te\'orica C-XI,
Universidad Aut\'onoma de Madrid, 
E-28049 Cantoblanco,
Madrid,
Spain;
e-mail: alejandro.saiz@uam.es, rosa.dominguez@uam.es}\hspace{0.3cm}
and
A. Serna
\footnote{Dpt. F\'{\i}sica y A.C., Universidad Miguel Hern\'andez,
E-03206 Elche, Alicante, Spain;
e-mail: arturo.serna@umh.es}}

\begin{abstract}
   We present   results of an  analysis
  of the structural and kinematical
properties of a sample of elliptical-like objects (ELOs)
 identified  in four hydrodynamical,   self-consistent
      simulations run with the DEVA code (Serna et al. 2003).
Star formation has been implemented in the code through a simple
phenomenological parameterization, that takes into account
stellar physics processes only implicitly through the values 
of a threshold gas density, $\rho_{\rm g,thres}$,
and an efficiency parameter, $c_*$.
  The four simulations operate in the context of a
$\Lambda$CDM cosmological model consistent with observations,
resolve ELO mass assembly at scales up to $\simeq$ 2 kpc,
and differ in the values of their star formation parameters.
 Stellar masses,
projected half-mass radii and central l.o.s. velocity dispersions,
$\sigma_{\rm los, 0}$, have been measured on the ELO sample
and their  values compared with data 
from the Sloan digital sky survey.
For the first time in self-consistent simulations,
a good degree of agreement has been shown, including
the Faber-Jackson  and  the $D_n - \sigma_{\rm los, 0}$ relations,
among others, but only 
when particular values of the  $\rho_{\rm g,thres}$ and $c_*$ parameters
are used. This demostrates the effect that the star formation parameterization
has on  the ELO mass distribution.
Additionally, our results suggest that  it is not strictly necessary,
at the scales resolved in this work,
to appeal to energy sources other than gravitational
(as for example   supernovae  feedback effects) to
account for the structure and kinematics of large ellipticals.

\end{abstract}


\section{INTRODUCTION}

One of the most challenging open problems in modern cosmology
is the origin of
the local galaxies of different Hubble types  we observe to-day.
Among them,  ellipticals
are the easiest to study.   
They form the  most homogeneous 
family and show the most precise regularities in the 
form of correlations among some pairs of their observable parameters.
The Sloan digital sky survey (SDSS, see York et al. 2000)
sample of early-type galaxies,
containing to date 9000 galaxies from different environments,
provides a new standard of 
reference for nearby elliptical  galaxies.
Its analysis confirmed correlations previously established,
such as those involving  structural and kinematical parameters
(the $L - \sigma_{\rm los, 0}$ or Faber-Jackson relation,
1976; the surface-brightness - effective radius relation,
Kormendy 1977; the $D_n - \sigma_{\rm los, 0}$ relation, Dressler et al. 1987;
among others, see Bernardi et al. 2003a, 2003b, 2003c, 2003d,
  and references quoted
therein).
These  correlations, as well as the [$\alpha/$Fe] ratio trend 
with $\sigma_{\rm los, 0}$ (Jorgensen 1999), 
demand  short formation time-scales and old formation
ages for the bulk of the stellar populations of ellipticals.
These requirements are naturally met by one of the scenarios
proposed so-far to explain galaxy formation and evolution:
the so-called {\it monolithic collapse} scenario (ellipticals
would form  at high $z$ in a single burst of star formation,
and would passively evolve since then; Patridge \& Peebles
1967; Larson 1974). 
The competing {\it merging  scenario} 
(galaxy mass assembly takes place gradually through repeated
mergers of smaller subunits; Toomre 1977; Kauffmann 1996)
meets some difficulties at
explaining the correlations above as well as other observations
on ellipticals, see Peebles 2002
 and Matteucci 2003 for  details and discussions.
 But the monolithic collapse scenario
does not recover all the currently available observations on ellipticals 
either.
 Such are, for example,  the range in ages their
 stellar populations
span in some cases or  their kinematical and dynamical peculiarities
 (Trager et al. 2000;
Menanteau, Abraham \& Ellis 2001), indicating
that
 an important fraction of
present-day ellipticals have recently experienced merger
events.

In order to reconcile all this observational background
within a formation scenario, it is preferable to study
galaxy assembly from simple physical principles and in
connection with the global cosmological model.
Self-consistent gravo-hydrodynamical simulations are a very
convenient tool to work out this problem 
(Navarro \& White 1994,
Tissera, Lambas \& Abadi 1997, Thacker \& Couchman 2000).
The method works as follows:
initial conditions are set at high
$z$ as a Montecarlo
realization of the field of primordial fluctuations
to a given cosmological model in a periodic, homogeneously sampled box;
 then the evolution of these fluctuations is numerically followed
up to $z =0$ by means of a computing code that solves the
N-body plus hydrodynamical
evolution equations.
In this way, the uncertainties resulting from prescriptions on
dynamics and gas  cooling and heating, present in other
methods such as semi-analytical ones (Kauffmann et al. 1999,
Mathis et al. 2002) can be removed, only star formation needs
further modelling. 
Individual galaxy-like objects (GLOs) naturally appear as an output
of the simulations,  
 no  prescriptions  are needed
as far as their mass  assembly processes 
are concerned. Moreover,
self-consistent simulations directly provide detailed information
on each individual GLO  at each
$ z$, namely  its  six dimensional phase space structure,
 as well as  the  temperature and age distributions of
its gaseous and stellar components, respectively.
 From this information, the 
 parameters  characterizing  each GLO
  can be estimated and compared
with observations (see, for example, S\'aiz et al. 2001,
concerning disk galaxies).
The first step in the program of studying the origin of galaxies
through self-consistent simulations, is to make sure that they
form GLO samples of different Hubble types that have counterparts
in the real local Universe. 
In particular, the possible simple correlations  involving  structural
and kinematical  parameters must be recovered.
 A detailed  analysis of this kind was
not yet available for ellipticals 
 (see previous work in Kobayashi 2002;
Sommer-Larsen, Gotz,  \& Portinari 2002;
Meza et al. 2003). 

We present in this paper the results of an analysis of
the structure and kinematics of a sample of
 elliptical-like objects (ELOs)
identified in four  self-consistent hydrodynamical simulations
run in the framework of a flat $\Lambda$CDM model consistent
with observations.
We have used 
DEVA,
an  AP3M-SPH code particularly designed to study
galaxy assembly in a cosmological context.
In DEVA, special attention has been paid that
the implementation of conservation laws (energy, entropy
and angular momentum) is as accurate as possible 
 (see Serna, Dom{\'{\i}}nguez-Tenreiro, \& S\'aiz, 2003 for details,
in particular for  a discussion on the observational implications of
violating some conservation laws).
Star formation (SF) processes have been implemented in the code
through a simple  parameterization,
similar to that first used by Katz (1992),  
that includes a  threshold gas density,
$\rho_{\rm g,thres}$ and an efficiency parameter, $c_{*}$, 
determining  the SF timescales
 according with a
Kennicutt-Schmidt law\footnote{See Elmegreen (2003) for a discussion
on the possibility that this law can be explained as a result of SF 
processes at the scale of molecular cloud cores, through an
interstellar medium  (ISM) gas structure whose density,
prior to SF, can be 
described by a log-normal probability distribution, as Wada \& Norman
(2001) found in their simulations
} (Kennicutt 1998).

Galaxy-like objects of different morphologies have been
 identified in the simulations 
(S\'aiz, Dom\'{\i}nguez-Tenreiro \& Serna 2002; S\'aiz 2003). 
The aim of this paper is to show that some of  those identified as 
 ELOs, have, at a structural and kinematical level,
counterparts in the local Universe, including parameter  correlations.
Data have been taken from the SDSS
 as analyzed by
Bernardi  et al.\ (2003a, 2003b),
Kauffmann et al. (2003a, 2003b) and
 Shen et al., 2003.
A brief account on ELO assembly is as follows: it mainly  occurs through
a multiclump  collapse at rather high $z$ 
  involving
many clumps; collapse 
 takes the clumps closer and closer along
filaments causing them to merge at very low relative angular
momentum and, consequently, at short timescales. This results
into  fast
SF bursts at high $z$ that transform most of the available gas
into stars.
The frequency of  head-on  mergers
 decreases with
$z$. ELO stellar populations are mostly old, and a trend exists 
with $\sigma_{\rm los, 0}$, as suggested by 
some observations (Thomas, Maraston \& Bender, 2002).     

\section{OBJECTS AT $z=0$}

We have analyzed ELOs identified in four different simulations,
namely S14, S16, S17 and S26, run within the same
global flat $\Lambda$ cosmological model, with $h=0.65$,
$\Omega_{\rm m} = 0.35$, $\Omega_{\rm b} = 0.06$.
The normalization parameter has been taken slightly high,
$\sigma_8 = 1.18$, as compared with the average fluctuations 
of 2dFGRS or SDSS  galaxies
(Lahav et al. 2002, Tegmark et al. 2003)
to mimic an active region of the Universe
(Evrard, Silk \& Szalay 1990).
The gravitational resolution is $\epsilon_{\rm g}$= 2.3 kpc.
Feedback effects from  stellar processes have  not been
explicitly included in the simulations, but the values of the
SF parameters we use mimic them in a sense
\footnote{Note that the role of stellar processes at kpc scales
is not yet clear,
as some authors argue that their  energy release drives
the structure of the ISM locally
at sub-kpc scales, while gas compression processes other
than stellar pressures, for example gravitational instabilities,
structure the gas prior to SF at scales $\ge$ kpc
(Elmegreen, 2003). Our ignorance of  sub-kpc (that is, subresolution)
scale processes, including SNe feedback effects, is circumvented 
by taking them implicitely  into
account  through the SF parameterization}.
S14, S16, and S26 share the same initial conditions 
(their  assembly histories are identical at halo scales)
 and differ only in the SF parameters
($c_*$ = 0.1, 0.03 and 0.01, and $\rho_{\rm g,thres}$ = $6 \times 10^{-25},
1.8 \times 10^{-24},6 \times 10^{-24}$ gr cm$^{-3}$ 
for S14, S16 and S26, respectively,
that is, SF becomes increasingly more difficult from S14 to S26), to test
the effects of SF parameterization to shape ELOs.
S17 is identical to S16, except that their initial conditions
differ, to test cosmic variance.
A standard cooling function has been used.
Each  of the four simulations started at a redshift $z_{\rm in} = 20$.
In any run, 64$^3$ dark matter and 64$^3$ baryon  particles,
with a mass of $1.29 \times 
10^8$ and $2.67 \times 10^7 $M$_{\odot}$, respectively, have been used
to  homogeneously sample the density field in a periodic box of 10 Mpc side.
The mass function of galaxy-like objects formed in this box is consistent
with that of a small group environment
(Cuesta-Bolao \& Serna, private communication),
 where the efficiency of galaxy
formation per volume unit is higher than average in the universe, 
and, so, the fraction of cold baryons (i.e., cold gas and stars)
in the box is also higher than average.
ELOs have been identified as those objects having a prominent stellar
spheroidal component
 with hardly disks at all
(S\'aiz 2003; S\'aiz et al., in preparation). 
The 8 more massive objects identified at $z=0$
in S14, S16 and S17, and the 4 more massive in S26, fulfill this 
condition. The most (least) massive object in the sample
has, at $z=0$,
 59,300 (4,130) dark and 29,600 (2,610) baryon particles
within its virial radius. The   spin parameters of the ELO sample have 
an average value of $\bar{\lambda} = 0.033$. 
Their stellar components have ellipsoidal shapes, and, in most cases,
they are  dynamically relaxed systems.

The 3D stellar mass density profiles are steeper than those of dark matter (DM)
as a result
of  energy losses by gas particles as they cool and fall onto
the center of the configuration.
The size scales are closely correlated to the halo mass scale, 
$M_{\rm vir}$; they
also depend on the SF parameters,  because the easier it is
for a gas particle in a given proto-ELO  to be transformed into a
stellar particle, the lower the amount
 of energy it will radiate before being 
transformed, and, so, the lower the total dissipation 
ensuing the ELO formation.
Concerning ELO velocity distributions, their 3D structure
is such that stars and dark matter do not exhibit a global rotation;
the stellar velocity dispersion profiles are systematically lower 
than DM ones (as found by Loewenstein 2000 on theoretical grounds)
and they slightly decrease outwards;
the anisotropy parameter 
 is always positive,
that is, disordered energy lies preferentially on radial motions. 
Projected  mass and velocity distributions  
can be characterized by global
mass (i.e., luminosity), size and velocity dispersion
parameters that have been extensively observed in real 
ellipticals  and can be easily measured in simulated ELOs,
making the comparison  with observational data possible.
Physically, the mass parameter at the ELO  scale
is $M_{\rm bo}^{\rm cb}$, 
the total amount of cold baryons that have reached
the central $\sim $ few kpc volume of the halos,
forming an ELO.    
A fraction of these cold baryons
have turned into stars, depending on
the strength of the dynamical  activity in the
volume surrounding the  proto-ELO
at high $z$, and, also, on the efficiency of the SF algorithm. 
$M_{\rm bo}^{\rm star}$ is the stellar mass. 
Hereafter we will only refer to $M_{\rm bo}^{\rm star}$ as mass
scale, as it can be estimated from    
 luminosity data  through modelling (Kauffmann et al.
2003a, for example).
Note that S14 objects have a slightly higher
stellar content 
than S26 objects, with those from S16 and S17 at intermediate
positions, as expected due to the values of their  SF parameters.
The projected  effective \emph{mass} radius,
$R_{\rm e, bo}^{\rm star}$,  is the radius
  enclosing a mass
 equal to  $M_{\rm bo}^{\rm star}/2$ as seen in projection.
Observationally,
a useful characterization of the
velocity dispersion of an elliptical galaxy is
provided by its central
line-of-sight velocity dispersion, $\sigma_{\rm los, 0}$. It
 has been measured in the ELO sample and found to be tightly
correlated with $M_{\rm vir}$.

Our next concern is related with the extent to which
ELOs in the sample as described by 
$M_{\rm bo}^{\rm star}$ as mass parameter,
 $R_{\rm e,bo}^{\rm star}$ as
projected size parameter, and $\sigma_{\rm los, 0}$
as velocity dispersion parameter, have observational
counterparts.      
On the observational side, Bernardi et al. (2003b)
provide maximum-likelihood estimates of the
parameters characterizing the joint probability distribution
(a trivariate gaussian) for  absolute luminosities, $M$,
(the logarithms of) intrinsic sizes, $R$, and central velocity
dispersions, $V$, namely, their mean values and the covariance
matrix.
Stellar masses of a sample of 10$^5$ SDSS galaxies
of different morphological types have been estimated by
Kauffmann et al. (2003a). Their results indicate that the
 stellar-mass-to-light ratio, $S$, can be taken to be constant in the
range of absolute luminosities $M < -21$, that is, for the more massive
or, equivalently, early-type galaxies in their sample
(see Kauffmann et al. 2003b).
The values of the logarithm 
of this ratio are  $S \simeq 0.53$ and $S \simeq 0.25$,
with dispersions $\sigma_S < 0.15$ and 0.1,  in the
$r$ and $z$ SDSS bands, respectively.
The independence of the value of $S$ with $M$ implies that
the pairwise correlation coefficients, $\rho_{\rm SM}$, can 
be taken $\simeq 0$ to a good accuracy.
Assuming, moreover, that the pairwise 
correlation coefficients $\rho_{\rm SR}$ and $   \rho_{\rm SV}$,
are also $\simeq 0$,
the pairwise concentration ellipses (and regression lines) can be drawn
for  the same data variables as those we have measured
on simulated ELOs (that is, stellar masses instead of absolute
luminosity, $R$ and $V$). This allows us to do  a more meaningful comparison.

\begin{figure}
\centerline{\includegraphics[height=15cm]{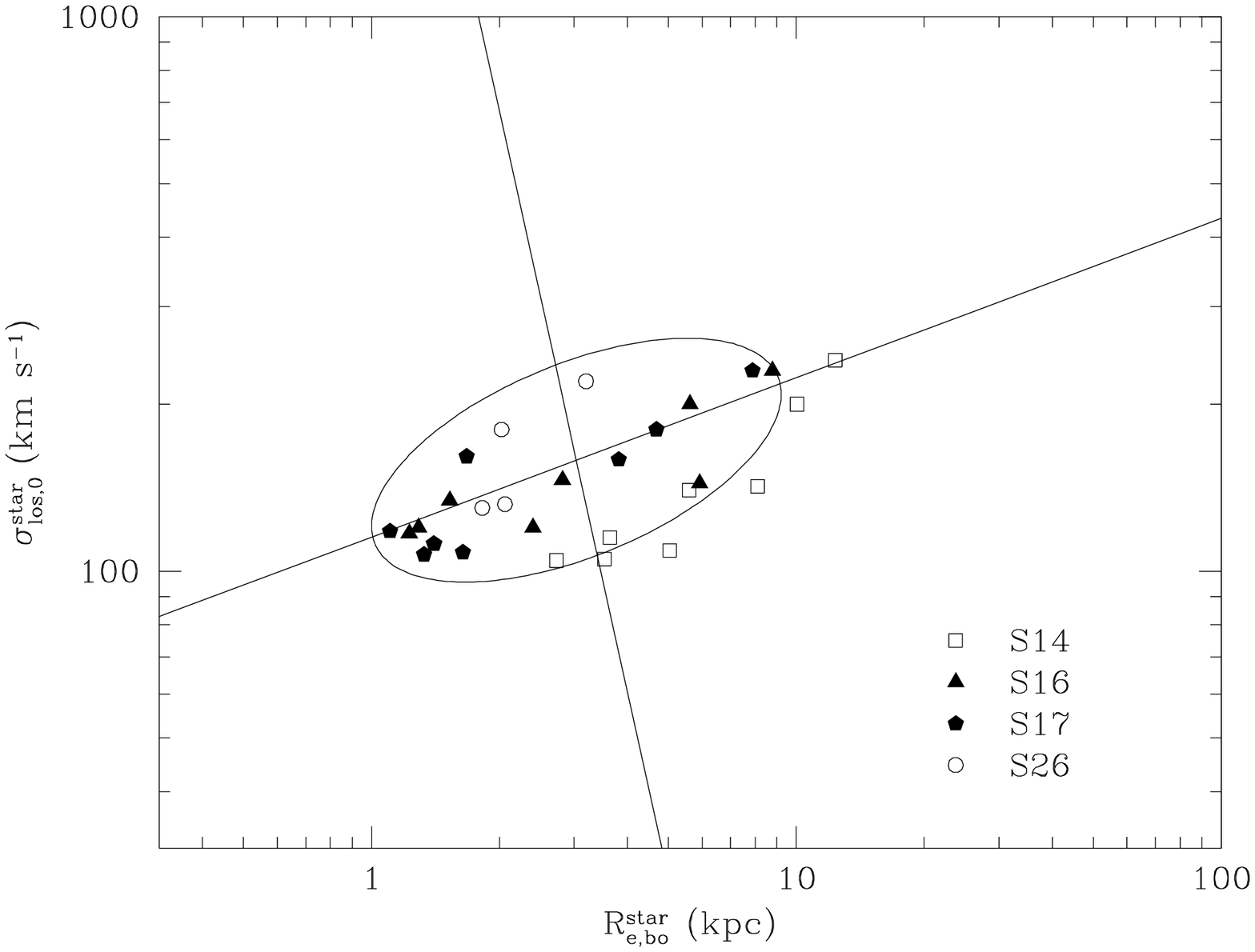}}
\caption[]{
The central l.o.s. velocity dispersions of  ELOs in the   sample versus
their  stellar  half-mass radii. Different symbols stand for different
simulations.
We also draw the concentration ellipse (with their major and minor
axes) for the SDSS early-type galaxy sample from Bernardi et al. (2003b) in the $z$-band.
See text for more details 
}
\label{Fig1ObsLetter}
\end{figure}

In Figure~\ref{Fig1ObsLetter},  the consistency 
between SDSS data and ELOs as seen in the
 $\sigma_{\rm los, 0}$ versus $R_{\rm e, bo}^{\rm star}$ plot
is addressed.
Points are ELO measured values, with different symbols for
different simulations. The ellipse is the 2$\sigma$
concentration ellipse and the lines are its major
and minor axes for
the SDSS early-type galaxy  sample in the $z$ band (the other SDSS bands
lead to almost identical data diagrams).
We recall that the major axis corresponds to  the orthogonal mean square
regression line, o.m.s.r.l.,
for the two variables in the Figure, and that
the 1$\sigma$ (2$\sigma$) region from the o.m.s.r.l. can be drawn
by tracing its parallel lines through the middle (end) points of the
minor semiaxis; they have not been drawn in the Figure for the sake
of clarity.
We see that, except for 5 of the S14  ELOs,
the whole ELO sample lies within the 2$\sigma$ concentration
ellipse.
The effects of the different SF parameterizations are apparent
in this plot. We see that while those ELOs formed in simulations
where SF is easy (S14), tend to be too large for their
$\sigma_{\rm los, 0}$ and outside the 1$\sigma$ 
(or even the 2$\sigma$) o.m.s.r.l. region,
those formed in simulations where SF is difficult (S26), tend to be 
smaller for their velocity dispersion.
ELOs formed in S16 and S17 have intermediate sizes and agree with
SDSS data at 1$\sigma$ level, showing a power-law correlation
(the $D_n - \sigma_{\rm los, 0}$ relation). 
This behaviour with the SF parameterization is consistent with
the interpretation that the easier to form stars, the lower
the amount of energy lost by the gas before being turned into stars.

\begin{figure}
  \centerline{\includegraphics[height=15cm]{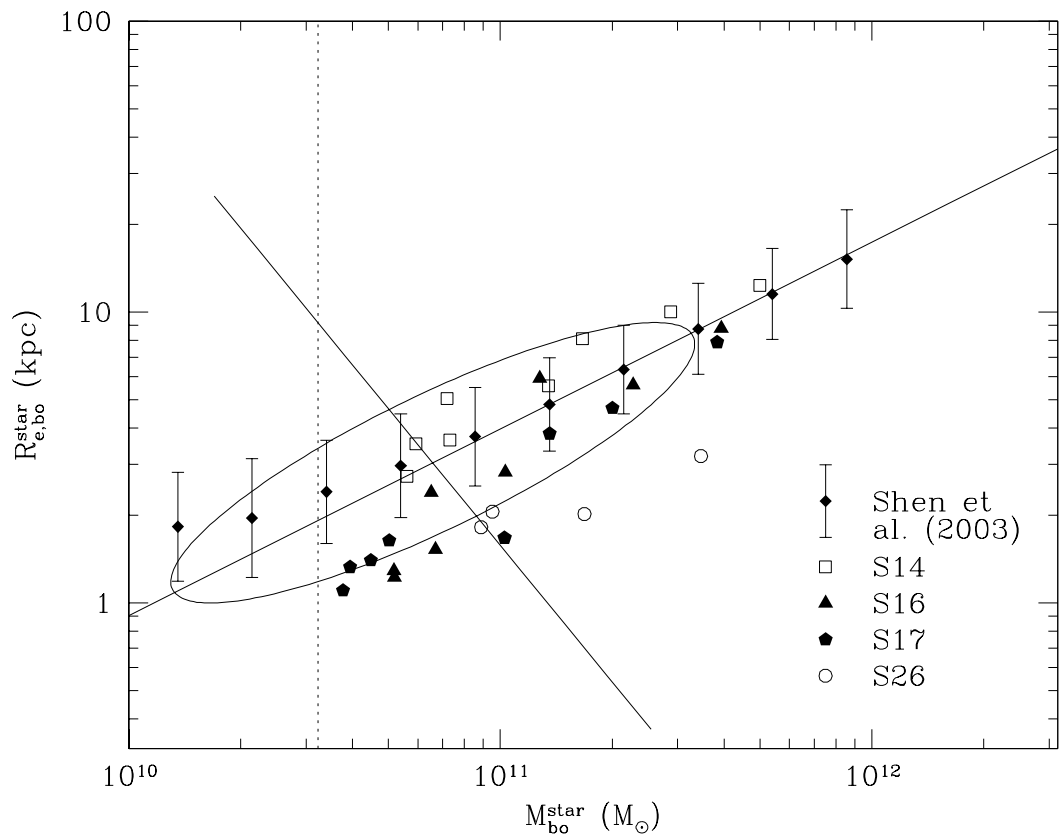}}
    \caption
      {
      The projected  stellar 
      half-mass radii versus 
      stellar masses.
    Filled diamonds with error bars correspond to
    median values of the S\'ersic half-light radii in the $z$-band
    and their 1 sigma dispersions (Shen et al. 2003).
We also draw the concentration ellipse 
for Bernardi et al.'s (2003b) sample in the $z$ band
and the lower limit for early-type-like galaxy stellar masses 
in the SDSS sample as found by
Kauffmann et al. (2003b, dotted vertical line).
See text for more details
      }
   \label{Fig2ObsLetter}
\end{figure}

In Figure~\ref{Fig2ObsLetter} we plot the stellar half-mass radii,
$R_{\rm e, bo}^{\rm star}$, versus the stellar masses,
$M_{\rm bo}^{\rm star}$.
Data are the median values (points) and
1 sigma dispersions (error bars) for the distribution of 
 S\'ersic (1968)  half-light radius in the $z$-band
as a function of stellar mass for SDSS early-type galaxies
(Shen et al. 2003, their Figure 11; data in the $r$-band do not
differ substantially).
We have also plot the 2$\sigma$ concentration ellipse drawn from
 Bernardi et al.'s
sample  as explained above.
The vertical line stands for
the  lower bound  of  early-type galaxy stellar masses,
  as found by Kauffmann
et al. 2003b.
We first note  that, when, say, 
$M_{\rm bo}^{\rm star} > 6 \times 10^{10} M_{\odot}$,
the consistency of data analysis by  Shen et al. 2003
and Bernardi et al. 2003 is remarkable.
Comparing Shen et al. 2003 results and our simulations
in this mass range, we see
that the agreement  
is good at a 1$\sigma$
level for S14 ELOs (even if they are larger than median 
 data values), as well as for the
  S17 and S16 ELO sample (smaller than median data values),
 except for three of them, that are  within  2$\sigma$ of Shen et al.'s
 data.
In this range of stellar mass, $R_{\rm e,bo}^{\rm star}$
and $M_{\rm bo}^{\rm star}$ show a power law correlation 
with a low dispersion and an exponent and
zero point consistent with
data. 
For lower masses, S17 and S16 ELOs are too small for their stellar
masses at
1$\sigma$, as are S26 ELOs in any range of stellar mass. 
The effects of the SF parameterization on this plot are similar
to those on Figure~\ref{Fig1ObsLetter} we have just discussed.

\begin{figure}
\centerline{\includegraphics[height=15cm]{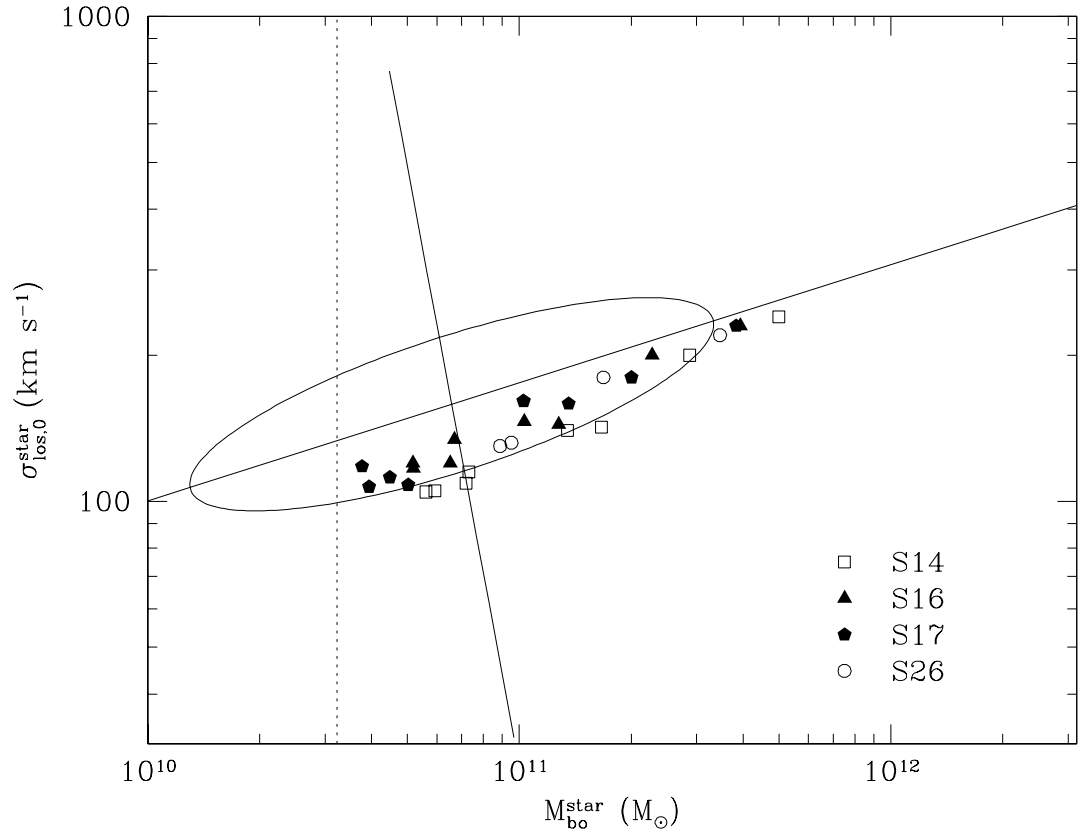}}
\caption{
   Same as  Figure 1 for the
   central l.o.s.\  velocity dispersions versus
   stellar masses. Vertical line is as in Figure 2
    }
\label{Fig3ObsLetter}
\end{figure}

Finally, we analyze the consistency between
simulations and data from the
$\sigma_{\rm los, 0}$ versus $M_{\rm bo}^{\rm star}$
plot (Figure~\ref{Fig3ObsLetter}).
We draw points for results of ELO measurements and
the 2$\sigma$ concentration ellipse for SDSS data. 
We see that most of the ELO sample lies within this  
ellipse, even if ELOs, and particularly so the less massive
ones, tend to have rather high stellar masses for their
$\sigma_{\rm los, 0}$ 
as compared with SDSS galaxies in the same $\sigma_{\rm los, 0}$
range.
Most ELOs with  $M_{\rm bo}^{\rm star} > 10^{11} M_{\odot}$,
 are inside the 1$\sigma$ o.m.s.r.l. region for SDSS data,
 and they show a power-law relation with slope
and zero point consistent with data and with low dispersion
(note that as $M^{\rm star}/L$ is constant, 
this is just the Faber-Jackson relation).
The dispersion is even lower for S17 and S16 ELOs with, say,
 $M_{\rm bo}^{\rm star} > 6 \times 10^{10} M_{\odot}$ when taken
 alone.
 Note that, even if most S14 objects
 have higher values of $M_{\rm bo}^{\rm star}$ than their
 S26 counterparts, with S16  and S17 objects at an intermediate position,
 all the
 objects follow the same relation, and particularly so the most massive ones,
  so that the different SF parameterizations
  have no important effects on this plot as far as the correlation
  is concerned.

\section{Discussion}

The degree of consistency reported here between sizes,
velocity dispersions and stellar masses of simulated
ELOs and SDSS data is very good. The agreement is
particularly outstanding when the simplicity of 
our working scheme is recalled: ELO assembly has been
simulated in the context of a cosmological model roughly
consistent with observations; Newton laws and hydrodynamical
equations have been integrated in this context,
with a standard cooling algorithm and a SF parameterization
through a Kennicutt-Schmidt-like law, 
containing our ignorance about its details at sub-kpc scales.
No further hypotheses to model  the assembly
 processes have been made.
Our results suggest that, at least  for the more massive objects,
say $\ge 6 \times 10^{10}$ M$_{\odot}$, it is not strictly
necessary to appeal to energy sources, {\it at the scales
resolved in this work}, that is, up to $\sim 2$ kpc, other than
gravitational to account for their sizes, velocity dispersions,
stellar masses and their correlations. 
This result is consistent with the idea that SNe explosions
are not relevant at these scales, but only at smaller scales,
to make shells and trigger star formation in molecular cloud cores.
Their effect would be accounted for in the SF parameters
(Elmegreen 2003).

This work strongly suggests that the SF parameterization
is a key ingredient to determine the 
compactness of elliptical galaxies.  
A good agreement has been found in the correlations
addressed here with SDSS data, but 
in the case of those involving sizes, only when  particular  values
of the SF  parameters are used. Our results 
 push the problem of elliptical 
galaxy formation from understanding 
their mass  assembly  at scales $>$  kpc in 
a cosmological context (Dom\'{\i}nguez-Tenreiro et al. 2003, in preparation), 
to understanding how SF was regulated
at high $z$ as the gas falls within
collapsing volumes, or other shock locations, so as to proceed 
 just with  the   efficiency necessary to  produce
the sizes observed in  to-day ellipticals.   

It is a pleasure to thank J. Silk and J. Sommer-Larsen for
useful information on the topics addressed in this paper.
This work was partially supported by the MCyT (Spain) through grant
AYA-0973 from the Programa Nacional de Astronom\'{\i}a 
y Astrof\'{\i}sica. We also thank the Centro de Computaci\'on
Cient\'{\i}fica (UAM, Spain) for computing facilities.

\clearpage

\clearpage

\clearpage

\end{document}